\documentclass[preprint,final,5p,times,twocolumn]{elsarticle}
\usepackage{amssymb}
\usepackage{graphicx}
\usepackage{multirow}
\usepackage{color}
\usepackage{amsthm}
\usepackage{amsmath}
\usepackage{widetext}
\usepackage{amssymb}
\usepackage{float}
\usepackage[utf8x]{inputenc}
\usepackage[numbers]{natbib}
\usepackage[colorlinks=true,citecolor=blue]{hyperref}
\journal{Physics Letters B}
\begin{document}
\begin{frontmatter}
\title{Core or Halo? Two-Fluid Analysis of Dark Matter-Admixed Quarkyonic Stars in the Multi-Messenger Era}
\author{Jeet Amrit Pattnaik$^{a*}$}
\ead{*Corresponding author,jeetamritboudh@gmail.com}
 \author{D. Dey$^{b,c}$}
 \author{M. Bhuyan$^{b}$}
 \author{R. N. Panda$^{a}$}
 \author{S. K. Patra$^{a}$}
\address[1]{Department of Physics, Siksha $'O'$ Anusandhan, Deemed to be University, Bhubaneswar-751030, India}
 \address[2]{Institute of Physics, Sachivalya Marg, Bhubaneswar-751005, India}
\address[3]{Homi Bhabha National Institute, Training School Complex, 
 Anushakti Nagar, Mumbai 400094, India}

\begin{abstract}
For the first time, we explore dark matter (DM) admixed quarkyonic stars (DAQSs) within a two-fluid formalism, where the normal/visible sector is modeled by a quarkyonic equation of state (EOS) in the Effective Relativistic Mean Field (E-RMF) framework and the DM component is treated as a degenerate fermionic gas with scalar and vector self-interactions. Our analysis begins with the mass–radius (M-R) relation, showing that the inclusion of DM enables stellar configurations to reach the mass range compatible with the GW190814 event. We identify both DM core and DM halo morphologies among the viable EOSs, with core dominated and halo dominated cases exhibiting distinct signatures. By fixing the stellar mass within the GW190814 range, we constrain the possible dark matter fractions and explore the role of different interaction channels. Using the EOSs consistent with these constraints, we further investigate the tidal deformability ($\Lambda$), moment of inertia (MOI), and stellar radius, finding broad agreement with constraints from GW170817, GW190814, and NICER. Finally, we compile the characteristic properties of DAQSs, including EOS type, DM fractions, morphology (core vs halo), and macroscopic observables in a comparative summary. This study provides a unified two-fluid framework to explore dense QCD matter and dark matter in the multi-messenger era, suggesting that the GW190814 secondary object could plausibly be interpreted as either a DM core or a DM halo quarkyonic star.
\end{abstract}

\end{frontmatter}

\section{Introduction}
\label{sec1}
\noindent
Neutron stars provide a unique laboratory to understand matter at densities well above nuclear saturation. At several times the saturation density, hadronic models alone become insufficient, as quark degrees of freedom emerge. The quarkyonic phase offers an intermediate state where quarks coexist with nucleonic excitations, stiffening the equation of state and altering internal energy and pressure distributions, which influence macroscopic properties such as mass, radius, and tidal deformability \cite{dey2024}.

At the same time, a wealth of astrophysical and cosmological observations point toward the presence of dark matter, including galaxy rotation curves, velocity dispersions, cluster dynamics, gravitational lensing, and the cosmic microwave background \cite{iva23}. These findings imply that dark matter cannot be explained as baryonic material but instead represents a novel non-luminous component that interacts only weakly with standard model particles \cite{Lars2000}. Among proposed candidates, weakly interacting massive particles (WIMPs) are especially compelling, as they naturally reproduce the observed relic abundance and can be efficiently captured by neutron stars through scattering with nucleons. The high density of neutron star interiors facilitates energy loss and accumulation of dark matter, making these objects natural astrophysical probes of the dark sector \cite{Kouvaris_2008, Kouvaris_2010, Goldman_1989, G_ver_2014, dark_matter_4}. Depending on the assumed interaction channel, such systems are modeled either in a single-fluid approach \cite{DM3, E-RMF1, DM1, dey2025, pattnaik2025darkmattereffectscurvature}, where dark matter couples non-gravitationally with ordinary matter through Higgs' portal, or in a two-fluid framework \cite{Sandin_2009,Xiang_2014,PhysRevD.105.123034,pinku2025} where the two components interact solely via gravity.

Recent observational constraints highlight the need for such studies:
\begin{itemize}
\item Mass limits: PSR J0740+6620 ($2.08 \pm 0.07,M_\odot$) and PSR J0952–0607 ($2.35 \pm 0.17,M_\odot$) \cite{Cromartie_2020, Romani_2022}.
\item Radius bounds: radius of a $1.4,M_\odot$ neutron star $\lesssim 13.5$ km from GW170817 and NICER \cite{PhysRevLett.119.161101, PhysRevLett.121.091102, Riley_2019, Miller_2019}.
\item Heavy compact objects: the secondary of GW190814 ($2.50$–$2.67,M_\odot$) challenges conventional EOS models \cite{Abbott_2020}.
\end{itemize}
These constraints motivate the exploration of quark–hadron crossover and quarkyonic EOS, which provide additional stiffness in the core of neutron stars. Even so, the admixture of dark matter may be required to explain the heaviest observed compact objects \cite{dey2024, rather21a}. 

In this letter, we address the challenges of explaining the secondary massive component in the GW190814 event, constraining radii in line with current observations, and probing the microphysics of dense QCD matter together with the nature of DM. This is attained by combining quarkyonic matter with a gravitationally coupled dark matter sector in a two-fluid approach within the E-RMF formalism, which naturally reveals both DM core and DM halo quarkyonic star configurations.

\section{Theoretical Methods}

\paragraph{Visible sector-}
The visible component of the star is presented within the Effective Relativistic Mean Field (E-RMF) framework, where nucleons interact through exchange of mesons with nonlinear couplings \cite{G3,IOPB-I}. At low densities, the system behaves as purely baryonic, but at higher densities, quark degrees of freedom are gradually included following the quarkyonic construction \cite{PhysRevLett.122.122701,PhysRevD.102.023021}. In here, the nucleons occupy a thin shell close to the Fermi surface, in contrast, the low momentum states inside the Fermi sea are filled by up (u) and down (d) quarks. The transition between the two regimes is regulated by the onset density $n_t$, while the QCD confinement scale ($\Lambda_{\rm cs}$ determines the width of the nucleonic shell.  

In the quarkyonic model, the total pressure and energy density are obtained by adding the contributions from the nucleonic shell and the quark Fermi sea, ensuring charge neutrality and $\beta$-equilibrium at each step. The inclusion of quark degrees of freedom above $n_t$ enhances the pressure support of the EOS, producing a stiffer behavior than purely baryonic cases. This stiffening behaviour enables quarkyonic stars to sustain larger maximum masses and radii consistent with present multimessenger observations. For full detailed formalism, one can refer to Ref. \cite{dey2024}.

\paragraph{Dark sector-}
In here, the dark matter candidates is treated as a fermions. Its pressure and energy density follow from standard Fermi integrals and are modified by self-interactions. The energy density and pressure are then written as, 
\begin{equation}
\mathcal{E}_{\text{DM}} = \frac{1}{\pi^2} \int_0^{k_D} dk \, k^2 \sqrt{k^2 + M_D^{*2}} 
+ \frac{g_{DV}^2}{2m_{DV}^2} \rho_D^2 
+ \frac{m_{DS}^2}{2g_{DS}^2} (M_D - M_D^*)^2,
\end{equation}

\begin{equation}
P_{\text{DM}} = \frac{1}{3\pi^2} \int_0^{k_D} dk \, \frac{k^4}{\sqrt{k^2 + M_D^{*2}}} 
+ \frac{g_{DV}^2}{2m_{DV}^2} \rho_D^2 
- \frac{m_{DS}^2}{2g_{DS}^2} (M_D - M_D^*)^2.
\end{equation}

The number density of dark matter (DM) is represented by $\rho_{D}$, while the effective mass of the DM candidate is expressed as 
$M_{D}^{*} = M_{D} - g_{DS}\phi_{D}$. The interaction strengths are characterized by two couplings: the attractive component $C_{DS} = g_{DS}/m_{DS}$ and the repulsive component $C_{DV} = g_{DV}/m_{DV}$, both defined in units of GeV$^{-1}$. 

\paragraph{Two-Fluid Approach:}
We adopt a two-fluid description in which normal/visible matter (NM) and dark matter (DM) are separately conserved but coupled through the gravitational field. The total stress-energy tensor is simply the sum of the two contributions,
\[
T^{\mu\nu} = T^{\mu\nu}_{(NM)} + T^{\mu\nu}_{(DM)} ,
\]
and the stellar equilibrium equations follow from solving the two-fluid 
Tolman-Oppenheimer-Volkoff (TOV) system \cite{Xiang_2014}. 
From these solutions, we compute the global observables such as the mass-radius relation, dimensionless tidal deformability $\Lambda$, and the total moment of inertia $I$ (in units of $10^{45}\,\mathrm{g\,cm^2}$) using the slow-rotation formalism \cite{hartle1967}. In the TFDM framework, these parameters play a crucial role, since they govern the self-interactions of DM particles mediated primarily through gravity. They also regulate how DM couples to the dark scalar field ($\phi_{D}$) and the dark vector field ($V_{D}^{\mu}$). Furthermore, this two-fluid approach makes it possible to investigate whether DM accumulates as halos or forms compact cores within neutron stars, with the distribution being strongly influenced by the DM particle mass and the strength of its interactions. 


\section{Numerical Results}
\label{results} 
\begin{figure}
\includegraphics[width=1.0 \columnwidth]{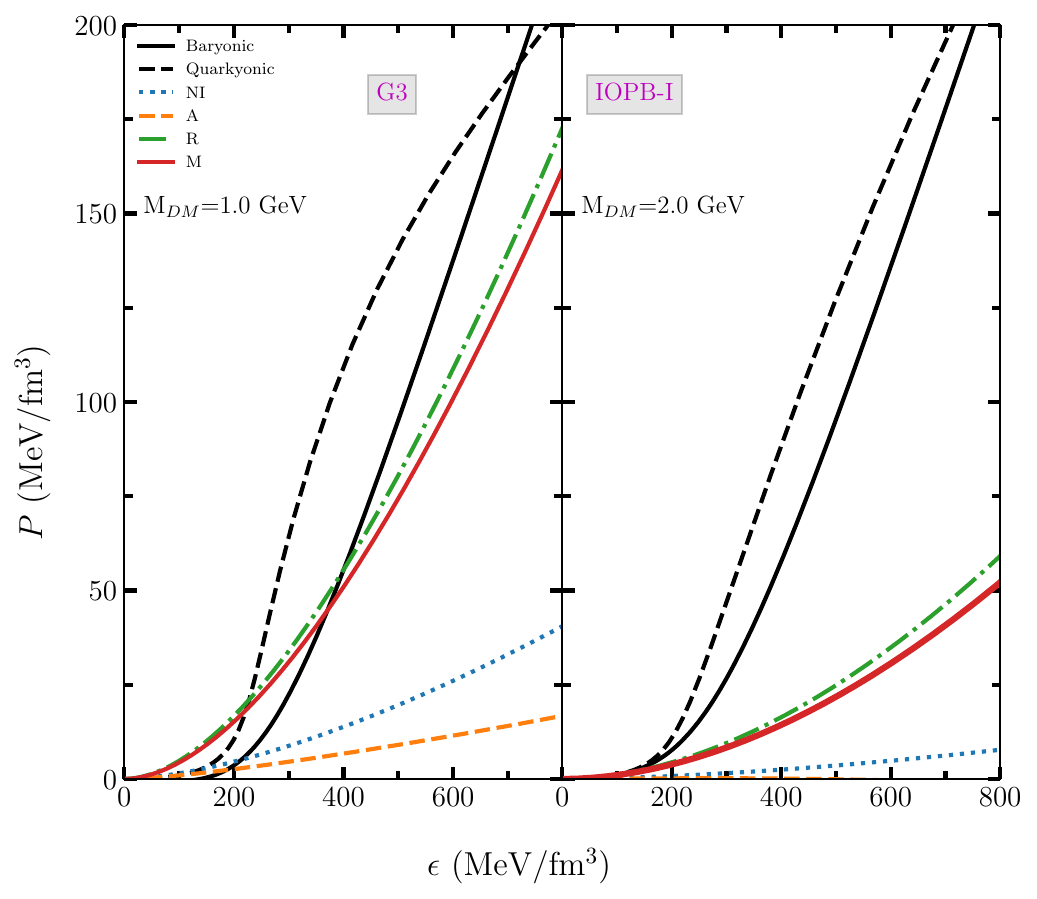}
\caption{{\it Left panel:} The EOS for normal/visible matter such as baryonic and quarkyonic matter ($n_t$= 0.3 $fm^{-3}$, $\Lambda_{\rm cs}$ = 800 MeV) for G3 force and DM EOS with $NI$ $\rightarrow$ No interaction (C$_{DV}$=0, C$_{DS}$=0), $A$ $\rightarrow$ Attraction (C$_{DV}$=0, C$_{DS}$=4 GeV$^{-1}$), $R$ $\rightarrow$ Repulsion (C$_{DV}$=10 GeV$^{-1}$, C$_{DS}$=0), $M$ $\rightarrow$ Mixed (C$_{DV}$=10 GeV$^{-1}$, C$_{DS}$=4 GeV$^{-1}$) at DM mass M$_{DM}$ = 1.0 GeV. {\it Right panel:} Same as left panel, but for IOPB-I force (normal/visible matter) and for various DM EOS at DM mass M$_{DM}$ = 2.0 GeV.}
\label{fig1}
\end{figure}

\begin{figure}
\includegraphics[width=1.0 \columnwidth]{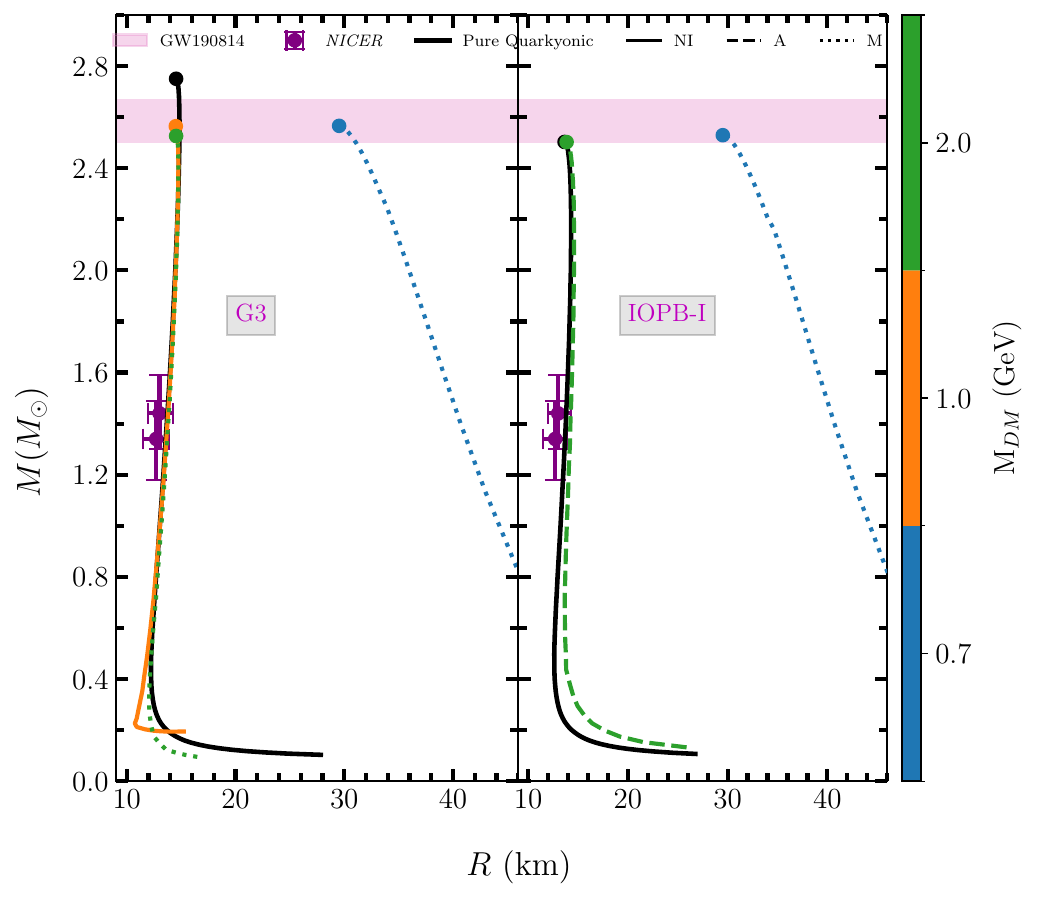}
\caption{The total M-R relation for pure quarkyonic ($n_t$= 0.3 $fm^{-3}$, $\Lambda_{\rm cs}$ = 800 MeV) and DM-admixed quarkyonic stars for various DM EOS with interaction type $NI$, $A$, and $M$. The colourbar represents the DM mass M$_{DM}$ = 0.7, 1.0, and 2.0 GeV. The solid dot refers to the Maximum mass (M$_\odot$). The observational GW190814 data and NICER radius measurements are also shown \cite{PhysRevLett.121.091102, Riley_2019, Miller_2019,Abbott_2020}.   }
\label{fig2}
\end{figure}

\begin{figure}
\includegraphics[width=1.0 \columnwidth]{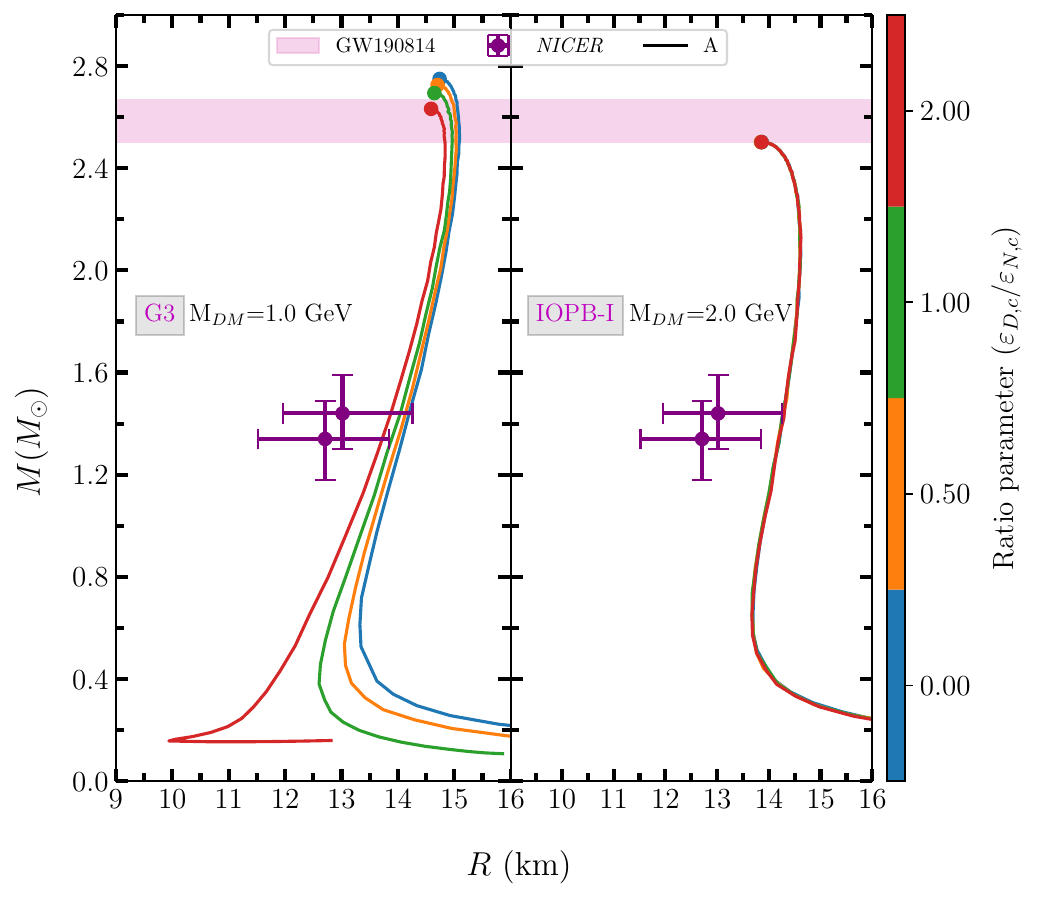}
\caption{The total M-R relation for DM-admixed quarkyonic stars for G3 and IOPB-I parameters considering $A$ (attraction) interaction type with DM mass M$_{DM}$ = 1 GeV and 2 GeV respectively. The colourbar represents the ratio parameter ($\frac{\epsilon_{D,c}}{\epsilon_{N,c}}$ ) ranging from 0.0 to 2.0.    }
\label{fig3}
\end{figure}

\begin{figure}
\includegraphics[width=1.0 \columnwidth]{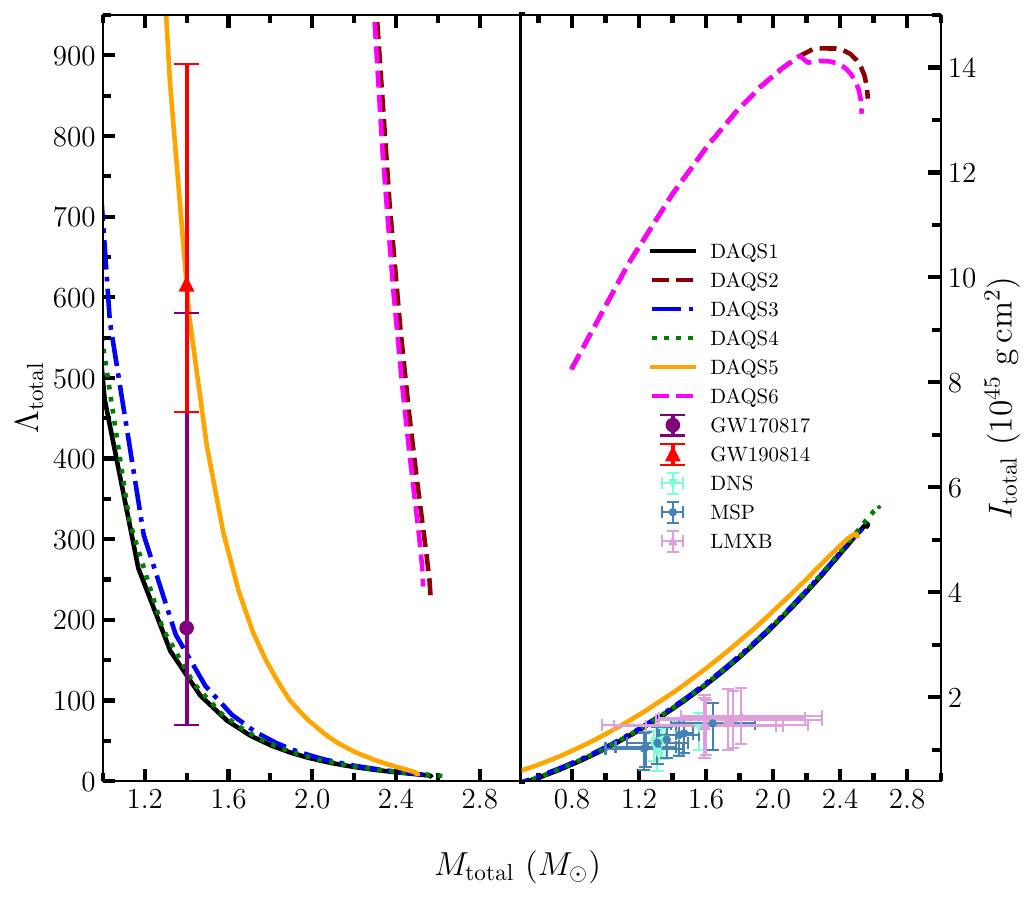}
\caption{{\it Left panel:} The total dimensionless tidal deformability ($\Lambda_{total}$) as a function of stellar mass M$_{total}$ (M$_{\odot}$) for EOS DAQS1-DAQS6. The purple and red colored error bars represent the observational constraints for events like the GW170817 NS-NS merger \cite{PhysRevLett.119.161101}, and the GW190814 BH-2.6 M$_{\odot}$ compact object merger \cite{Abbott_2020}). {\it Right panel:} The total moment of inertia (MOI) I$_{total} (10^{45} g cm^{2}$) of the NS as a function of total mass with the observational constraints \cite{Landry_2019}.  }
\label{fig4}
\end{figure}

\begin{figure}
\includegraphics[width=1.0 \columnwidth]{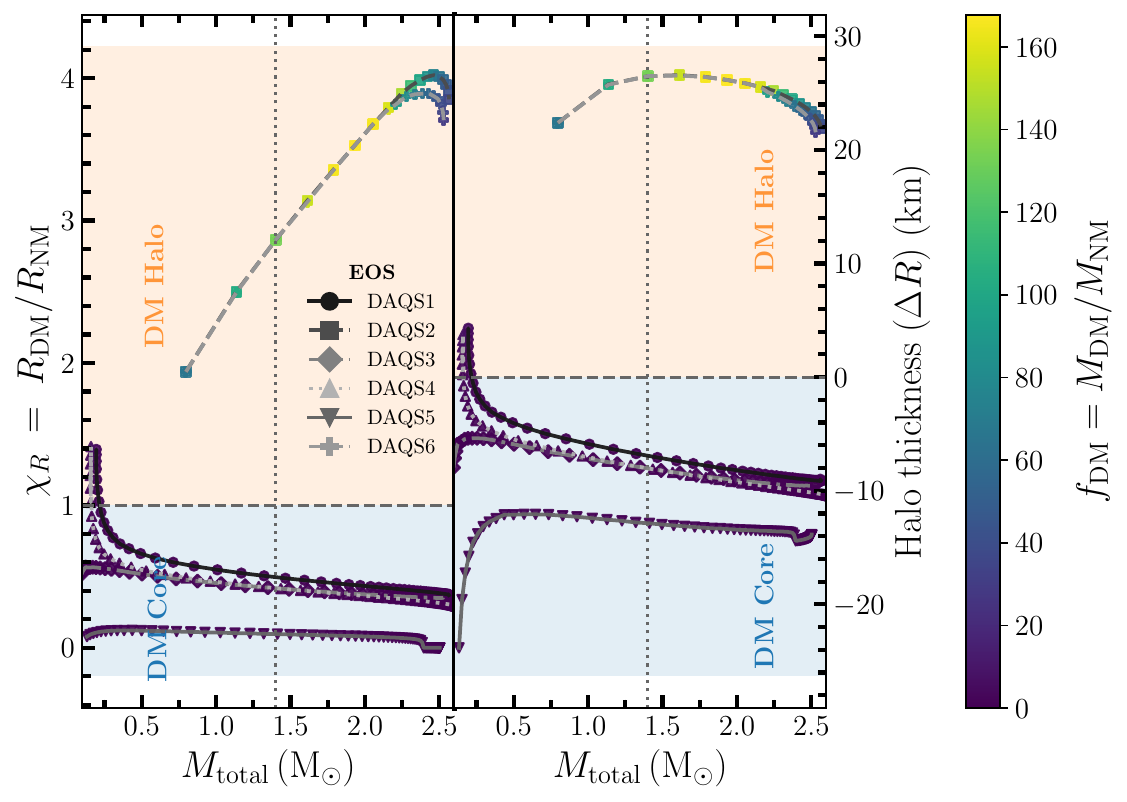}
\caption{The dark to normal matter radius ratio, $\chi_{R} = R_{\rm DM}/R_{\rm NM}$ and the halo thickness 
$\Delta R = R_{\rm DM} - R_{\rm NM}$ as function of total mass for DAQS1-DAQS6. The dashed line at $\chi_{R}=1$ and $\Delta R=0$ separates DM core ($\chi_{R}<1$, $\Delta R<0$ ) (blue band) and DM halo ($\chi_{R}>1$, $\Delta R>0$) (orange band) morphologies. The vertical line highlights the canonical $1.4\,M_{\odot}$ configuration. The color shading encodes the DM mass fraction $f_{\rm DM}$ where available.}
\label{fig5}
\end{figure}

\begin{figure}
\centering
\includegraphics[width=1 \columnwidth]{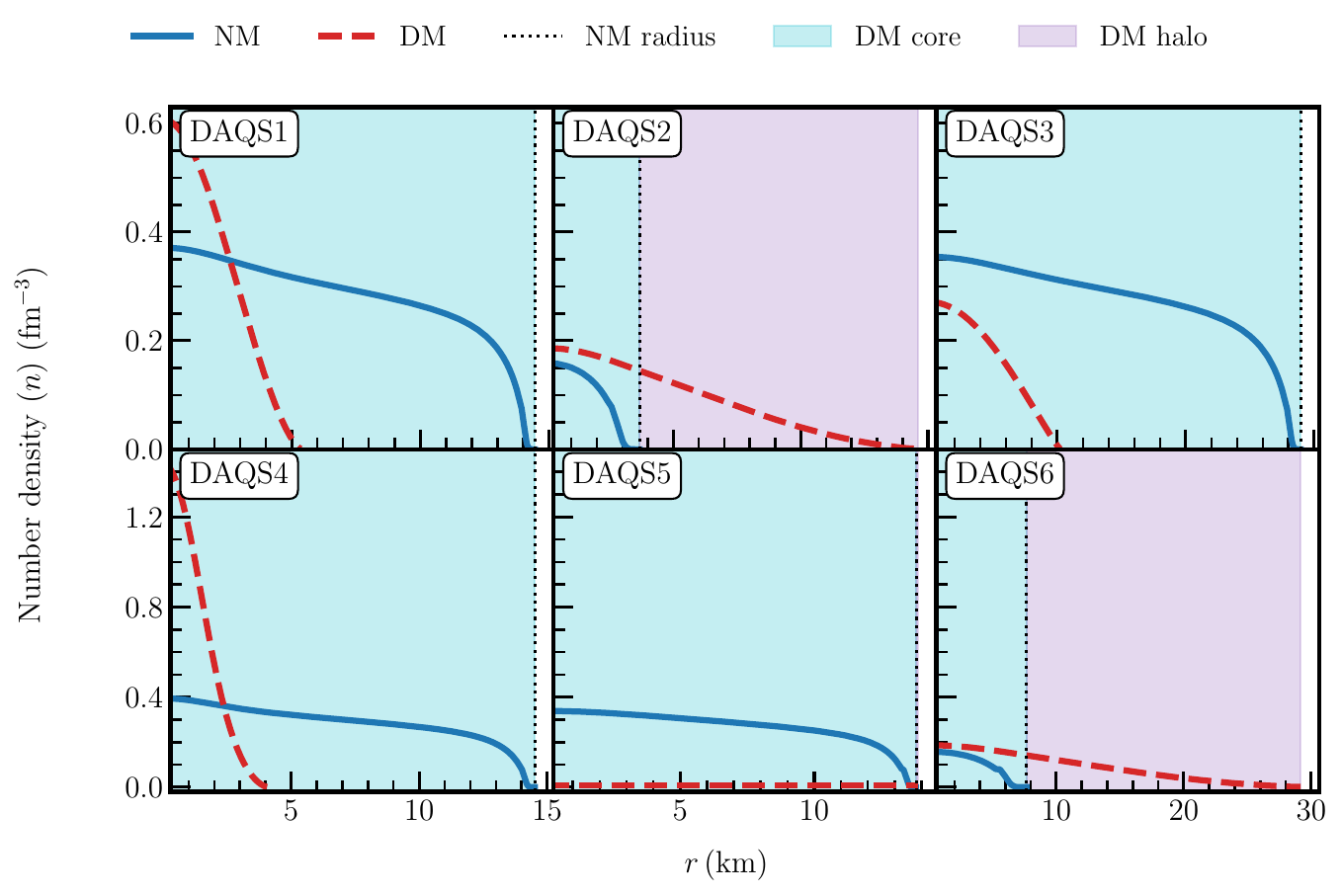}
\caption{Number density ($n$) (fm$^{-3}$) profiles of maximum mass dark matter-admixed quarkyonic stars for EOS DAQS1-DAQS6. The shaded regions represents the DM core and halo configurations. }
\label{fig6}
\end{figure}

We perform a systematic scan over DM admixed quarkyonic stars (DAQSs), focusing on the dark matter mass, fraction, and interaction type that can account for the available observational constraints. In this work, we concentrate on the primary quarkyonic star configuration characterized by a transition density of $n_t = 0.3~\text{fm}^{-3}$ and a QCD confinement scale of $\Lambda_{cs} = 800~\text{MeV}$ (See Fig. \ref{fig1}). The mass–radius (M-R) relations, dimensionless tidal deformability, moment of inertia are analyzed in detail in Figs. \ref{fig2}–\ref{fig6}. Finally, the resulting equation of state (EOS) characteristics are summarized in Table \ref{tab1}.\\

In Fig. \ref{fig1}, the visible sector, considering the baryonic equation of state, shows that the G3 force appears to be softer than IOPB-I, and this trend remains consistent even after the 
inclusion of quarkyonic matter. The stiffening of the quarkyonic EOS originates from the onset of additional quark degrees of freedom at high density, which enhances the pressure. In the dark sector, the ordering of stiffness is reversed depending on the interaction channel: a repulsive interaction ($C_{DV}=10 GeV^{-1}, C_{DS}=0$) yields the stiffest EOS, followed by the mixed repulsive-attractive case ($C_{DV}=10GeV^{-1}, C_{DS}=4GeV^{-1}$), then the non interacting scenario ($C_{DV}=0, C_{DS}=0$), and finally the purely attractive interaction ($C_{DV}=0, C_{DS}=4GeV^{-1}$), which produces the softest EOS. Further, the inclusion of DM with lower particle mass tends to stiffen the EOS more strongly. Notably, the EOS becomes softer for a DM particle mass of $2~\mathrm{GeV}$ compared to $1~\mathrm{GeV}$, indicating that heavier fermionic DM provides less degeneracy pressure and therefore contributes less effectively to the pressure support inside the star. This softening translates directly into smaller stellar radii and reduced tidal deformabilities. The same stiffness hierarchy is reflected in the mass-radius (M-R) relations in Fig. \ref{fig2} and Fig. \ref{fig3}.
%

In Fig. \ref{fig2}, a pure quarkyonic star ($n_t=0.3~\mathrm{fm^{-3}}$, $\Lambda_{\rm cs}=800~\mathrm{MeV}$) without dark matter yields a maximum mass of $2.75\,M_\odot$ and $2.50\,M_\odot$ for the G3 and IOPB-I forces, respectively. We systematically investigated all possible M-R relations by considering different DM particle masses and interaction types (no interaction, attractive, repulsive, and mixed). Among these, five suitable cases are identified that fall within the secondary mass range of GW190814. Interestingly, DM-admixed quarkyonic stars (DAQSs) with particle masses of 1 and 2 GeV exhibit broadly similar 
M-R patterns, whereas the case with $0.7~\mathrm{GeV}$ shows distinct deviations. The latter suggests the development of a DM halo configuration, leading to anomalous tidal deformabilities and unusually large moments of inertia, which place these models outside the NICER observational bounds. In contrast, the other cases remain consistent with both NICER and the tidal deformability constraints from GW170817. 

Further, we extended our analysis by introducing the ratio parameter  $\frac{\epsilon_{D,c}}{\epsilon_{N,c}},$ ranging from 0.0 to 2.0 (as shown in Fig. \ref{fig3}), where $\epsilon_{D,c}$ and $\epsilon_{N,c}$ denote the central energy densities of dark matter and normal matter, respectively. In this case, we focused on the attractive ($A$) interaction type DM EOS for both G3 and IOPB-I parametrizations. For the G3 force with $M_{\rm DM}=1.0~\mathrm{GeV}$, we observe that the M-R curve shifts towards smaller radii and the maximum mass decreases with increasing ratio parameter from 0.0 to 2.0. In contrast, for the IOPB-I EOS, no significant variation of the M-R profile is found once $M_{\rm DM}$ exceeds $2.0~\mathrm{GeV}$; the behavior essentially saturates. This saturation in IOPB-I arises because its intrinsic stiffness reduces the sensitivity of the M-R relation to additional DM fractions. From this analysis, we identify another possible candidate for DAQS consistent with the GW190814 secondary mass range: the G3 EOS with $M_{\rm DM}=1.0~\mathrm{GeV}$ 
at ratio parameter $=2.0$. 

\begin{table}[h!]
\centering
\renewcommand{\arraystretch}{1}
\setlength{\tabcolsep}{2pt}
\caption{EOS characteristics such as DM particle mass, interaction type, and ratio parameter $\epsilon_{D,c}/\epsilon_{N,c}$ are listed for the six DAQS configurations consistent with the GW190814 secondary mass range.}
\begin{tabular}{c c c c c}
\hline\hline
Star(s) & EOS   & $M_{\rm DM}$ (GeV) & DM Interaction Type & Ratio \\
\hline
DAQS1 & G3    & 1.0  & No Interaction        & 1.0 \\
DAQS2 & G3    & 0.7  & Mixed (Attr.+Repl.)  & 1.0 \\
DAQS3 & G3    & 2.0  & Mixed (Attr.+Repl.)  & 1.0 \\
DAQS4 & G3    & 1.0  & Attractive           & 2.0 \\
DAQS5 & IOPB-I  & 2.0  & Attractive           & 1.0 \\
DAQS6 & IOPB-I  & 0.7  & Mixed (Attr.+Repl.)  & 1.0 \\
\hline
\label{tab1}
\end{tabular}
\end{table}
After constraining the EOSs with the GW190814 secondary mass, we label them as DAQS1-DAQS6. Their defining characteristics, including the type of DM interaction, DM particle mass, and ratio parameter, are summarized in Table \ref{tab1}. Using these six EOSs, we compute the total tidal deformability $\Lambda$ and the total moment of inertia $I$ (in units of $10^{45}~\mathrm{g\,cm^2}$), which can be seen in Fig. \ref{fig4}. In general, $\Lambda$ is highly sensitive to the stellar radius ($\Lambda \propto R^5$) and therefore provides a stringent constraint on the EOS, while the moment of inertia offers an additional and independent probe of stellar structure.  

Among the six candidates, DAQS2 (G3, $M_{\rm DM}=0.7~\mathrm{GeV}$, mixed interaction, ratio = 1) and DAQS6 (IOPB-I, $M_{\rm DM}=0.7~\mathrm{GeV}$, mixed interaction, ratio = 1) display anomalous behavior. Their total tidal deformability remains almost constant with mass, producing a nearly flat profile rather than the expected smooth decrease. At the same time, their predicted total moments of inertia are abnormally large, reaching $I \simeq 8$-$15 \times 10^{45}~\mathrm{g\,cm^2}$, well above the typical interval of $0.4$-$6 \times 10^{45}~\mathrm{g\,cm^2}$ predicted by the other EOSs. When compared with astrophysical measurements of moment of inertia from double neutron stars (DNS), millisecond pulsars (MSP), and low-mass X-ray binaries (LMXB), only DAQS2 and DAQS6 fall outside the observational bands. This strongly suggests that these two cases may host extended DM halos, which redistribute mass into the outer layers, inflate $I$, and flatten the $\Lambda$-$M$ relation.  

In contrast, DAQS5 (IOPB-I, $M_{\rm DM}=2~\mathrm{GeV}$, attractive interaction, ratio = 1.0) is consistent with the GW190814 tidal deformability constraint. In addition, DAQS1 (G3, $M_{\rm DM}=1~\mathrm{GeV}$, no interaction, ratio = 1), DAQS3 (G3, $M_{\rm DM}=2~\mathrm{GeV}$, mixed interaction, ratio = 1), and DAQS4 (G3, 
$M_{\rm DM}=1~\mathrm{GeV}$, attractive interaction, ratio = 2) satisfy the tidal deformability bounds from GW170817. For the total moment of inertia, all EOSs remain compatible with DNS, MSP, and LMXB data, except for the anomalous DAQS2 and DAQS6.

Furthermore, to investigate the internal morphology of the DAQS configurations more precisely in Fig. \ref{fig5}, we analyse the radius ratio $\chi_R = \frac{R_{\rm DM}}{R_{\rm NM}},$ which distinguishes between DM core ($\chi_R < 1$) and DM halo ($\chi_R > 1$) structures. The six DAQS models identified in Table \ref{tab1} exhibit a variety of behaviors. DAQS2 and DAQS6, both with $M_{\rm DM} = 0.7~\mathrm{GeV}$ and mixed interactions, lie firmly in the DM halo regime, with $\chi_R > 1$ across the entire mass range. In these cases, $\chi_R$ increases with stellar mass, and its value is largest at the maximum mass, indicating that the halo becomes more extended for heavier configurations.  

In contrast, DAQS3 (G3, $M_{\rm DM}=2~\mathrm{GeV}$, mixed interaction, ratio = 1) and DAQS5 (IOPB-I, $M_{\rm DM}=2~\mathrm{GeV}$, attractive interaction, ratio = 1) remain pure DM core cases with $\chi_R < 1$ throughout, and the radius ratio decreases as the star approaches its maximum mass. The remaining two cases, DAQS1 and DAQS4, show a transitional behavior: at lower masses ($M \sim 0.2~M_\odot$) the stars develop light DM halos with $\chi_R > 1$, but as the mass increases toward $M_{\rm max}$, the DM retreats into the interior and the configurations evolve into DM cores ($\chi_R < 1$).  

A similar pattern is observed when examining the halo thickness 
$\Delta R = R_{\rm DM} - R_{\rm NM}.$ While $\chi_R$ is dimensionless and provides a qualitative insight, the $\Delta R$ carries dimensions (km) and directly measures the geometric extent of the halo. In DAQS2 and DAQS6, $\Delta R$ grows steadily with mass, reinforcing their halo nature. In DAQS3 and DAQS5, $\Delta R$ remains negative and decreases with mass, consistent with compact DM cores. For DAQS1 and DAQS4, $\Delta R$ changes sign, marking the halo-to-core transition with increasing stellar mass.  

The colour bar in Fig. \ref{fig5} encodes the DM mass fraction $f_{\rm DM}=M_{\rm DM}/M_{\rm NM}$, which further clarifies this classification. We find that the halo-dominated cases (DAQS2 and DAQS6) are associated with larger $f_{\rm DM}$ values, indicating that an extended DM distribution requires a relatively higher dark fraction. In contrast, the core-dominated cases (DAQS3 and DAQS5) correspond to smaller and more stable $f_{\rm DM}$, while the transitional models (DAQS1 and DAQS4) exhibit intermediate fractions that evolve with stellar mass. Taken together, $\chi_R$, $\Delta R$, and $f_{\rm DM}$ provide insight into whether DM forms a core, a halo, or a transitional configuration in quarkyonic stars.

We now examine the number density profiles of their maximum-mass configurations. Fig. \ref{fig6} shows the radial dependence of the normal/visible matter (NM) and dark matter (DM) densities for DAQS1-DAQS6. This analysis provides a direct visualization of whether the DM forms a compact core or an extended halo. In DAQS2 and DAQS6 ($M_{\rm DM}=0.7~\mathrm{GeV}$, mixed interaction), the densities of NM and DM start from nearly the same value at $r \to 0$, but the component of DM falls off more slowly and extends several kms beyond the NM surface, producing extended DM halos. This morphology explains the anomalous tidal and MOI behavior identified earlier. In contrast, the remaining four EOSs produce DM cores of varying strength. In DAQS1, the central DM density ($n_{\rm DM}\!\sim\!0.6$ fm$^{-3}$) is higher than that of NM ($n_{\rm NM}\!\sim\!0.4$ fm$^{-3}$), yielding a compact DM core. In DAQS4, this effect is even stronger, with the DM density peaking around $1.4$ fm$^{-3}$, far exceeding the NM density ($\sim 0.4$ fm$^{-3}$). DAQS3 shows NM dominance across the profile, with the DM distribution terminating near $r\simeq 10$ km while the NM extends to $\simeq 29$ km. DAQS5 represents the weakest DM case: the NM density exceeds the DM density, which remains almost flat and close to zero, producing a barely perceptible DM core.  
\section{Summary and Conclusions}\label{summary}
In this work, we have systematically explored quarkyonic stars admixed with dark matter within the two-fluid approach in the framework of E-RMF. By constraining six representative EOSs (DAQS1-DAQS6) against the maximum mass of GW190814, we classified their internal morphologies using a combination of tidal deformability, moment of inertia, radius ratio $\chi_R$, halo thickness $\Delta R$, and radial number density profiles. Our approach reveals a clear difference that DAQS2 and DAQS6 host extended DM halos, while DAQS1, DAQS3, DAQS4, and DAQS5 contain compact DM cores of varying strength.  

A particularly intriguing outcome is that both halo-dominated and core-dominated configurations can reproduce the GW190814 secondary mass, suggesting that this object could potentially be interpreted as either a DM halo star or a DM core star. The observational anomalies in tidal deformability and moment of inertia associated with halo cases highlight the possibility that multi-messenger data might distinguish between these two scenarios in the future. Thus, GW190814 offers a unique laboratory for testing dark matter imprints in compact stars, and our results demonstrate that quarkyonic matter with DM admixture provides a viable pathway to connect nuclear microphysics, dark sector properties, and astrophysical observations.

\bibliography{DAQS}
\bibliographystyle{elsarticle-num.bst}

\end{document}